\documentclass[floatfix, twocolumn]{revtex4}
\usepackage{graphicx,hyperref}

\begin{document}
\title{Low-frequency Noise in Josephson Junctions for Superconducting Qubits}
 \date{\today}
\author{J.~Eroms}\email{eroms@qt.tn.tudelft.nl}
\author{L.~C.~van~Schaarenburg}
\author{E.~F.~C.~Driessen}
\author{J.~H.~Plantenberg}
\author{C.~M.~Huizinga}
\author{R.~N.~Schouten}
\author{A.~H.~Verbruggen}
\author{C.~J.~P.~M.~Harmans}
\author{J.~E.~Mooij}
\affiliation{Kavli Institute of Nanoscience, Delft University of
Technology, Lorentzweg 1, 2628 CJ Delft, The Netherlands}
\begin{abstract}
We have studied low-frequency resistance fluctuations in
shadow-evaporated Al/AlO$_x$/Al tunnel junctions. Between 300~K and
5~K the spectral density follows a $1/f$-law. Below 5~K, individual
defects distort the $1/f$-shape of the spectrum. The spectral
density decreases linearly with temperature between 150~K and 1~K
and saturates below 0.8~K.  At 4.2~K, the spectral density is about
two orders of magnitude lower than expected from a recent survey
[D.~J.~Van~Harlingen {\em et al.}, Phys.\ Rev.~B~{\bf 70}, 064510
(2004)]. Due to the saturation below 0.8~K the estimated qubit
dephasing times at 100 mK are only about two times longer than
calculated by Van~Harlingen {\em et al.}
\end{abstract}
\maketitle

Superconducting qubits are promising candidates for a solid-state
realization of a quantum computer. Several groups have demonstrated
high-quality operation of single or coupled qubits working in the
charge~\cite{Bib:charge}, charge-phase~\cite{Bib:Collin},
flux~\cite{Bib:Irinel}, or phase regime~\cite{Bib:coupledPhase}.
Decoherence due to external sources such as the measurement devices,
has been studied extensively and is by now well
understood~\cite{Bib:Ithier}, permitting qubit dephasing times of up
to several microseconds~\cite{Bib:Patrice4us}. Future progress in
this field of research depends crucially on understanding and
controlling decoherence due to defects in the
devices~\cite{Bib:Astafiev,Bib:Simmonds,Bib:vH}. Superconducting
qubits contain Josephson junctions, whose Josephson energy $E_J =
\Phi_0 I_C/(2 \pi)$ determines the potential landscape of the qubit
($I_C$ is the critical current, and $\Phi_0=h/2e$ is the
superconducting flux quantum). Due to imperfections of the tunnel
barrier, $E_J$ fluctuates in time, leading to fluctuations in the
qubit potential. Therefore, the qubit energy splitting is not
constant during an experiment, which leads to decoherence.

The role of critical current noise in superconducting qubits was
addressed by Van Harlingen {\em et al.}~\cite{Bib:vH}. They
calculated the sensitivity to $I_C$-noise for different qubit
designs and estimated decoherence for various data acquisition
schemes, assuming a $1/f$-dependence of the noise spectral density.
For a quantitative comparison, they compiled $1/f$-noise strengths
from published data, which turned out to be remarkably universal at
4.2~K. To extrapolate to millikelvin temperatures where qubits are
operated, they assumed a $T^2$-scaling of the spectral density,
based on an experimental study~\cite{Bib:Wellstood}. This assumption
is supported by a recent theoretical paper~\cite{Bib:Shnirman}. %
The experimental data compiled in Ref.~\onlinecite{Bib:vH} did not
contain measurements from the Al/AlO$_x$/Al material system, which
is utilized by most groups working on superconducting qubits. Also,
no sub-micron junctions were included in the survey. For those
junctions, it was demonstrated that individual bistable defects in
the barrier can distort the $1/f$-shape of the noise spectral
density~\cite{Bib:Rogers}. Thus, for a reliable estimate of qubit
dephasing times, essential parameters were still lacking. This
motivated us to investigate low-frequency $I_C$-fluctuations in Al
junctions prepared by the double-angle shadow evaporation
method~\cite{Bib:Dolan}, which is employed for many qubit
experiments.
\begin{figure}\includegraphics[width=8.5cm]{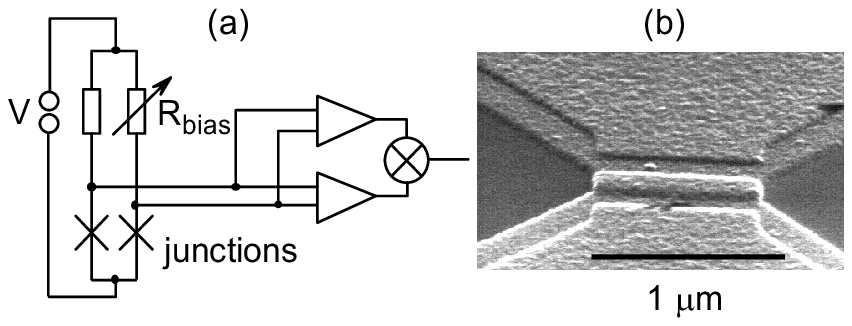}
\caption{(a) Simplified schematics of the measurement setup using a
resistance bridge circuit and cross correlation of the amplifier
signals. (b) Scanning electron micrograph of a shadow evaporated
tunnel junction, viewing angle 60$^\mathrm{o}$ to the surface
normal. The Al leads are widened immediately after the junction to
enhance cooling.} \label{Fig:Setup}
\end{figure}

Low-frequency noise of a Josephson junction is caused by bistable
defects situated in the tunnel barrier~\cite{Bib:Rogers}. The
fluctuating tunnel transparency results in fluctuations in the
tunnel resistance $R_N$ and the critical current $I_C$.  Each
individual defect produces a random telegraph signal in the device
resistance with a Lorentzian spectrum~\cite{Bib:DH,Bib:Rogers}
\begin{equation}
S_R(f) = \frac{\delta R^2 \tau_{\mathrm{eff}}}{1 + (2\pi f
\tau_{\mathrm{eff}})^2}\qquad ,\label{Eq:Lorentz}
\end{equation}
where $\delta R$ is the resistance change caused by the defect and
$\tau_{\mathrm{eff}}$ is the average transition rate between the two
states. If the random telegraph signals of many weak defects with a
range of $\delta R$ and $\tau_{\mathrm{eff}}$ are superimposed, the
total noise spectrum of the tunnel junction shows a $1/f$
shape~\cite{Bib:DH}. In sub-micron junctions however, a small number
of strong defects dominate the noise spectral density, giving rise
to one or more Lorentzians on top of a $1/f$
background~\cite{Bib:Rogers}.

\begin{figure}\includegraphics[width=8.5cm]{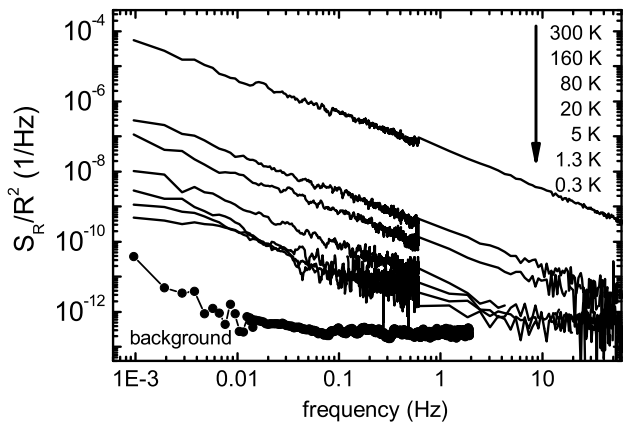}
\caption{Noise spectral density of a sample with
$880\times190$~nm$^2$ junction area from room temperature (topmost
trace) to 0.3~K (lowest trace). The data is normalized to the tunnel
resistance of the sample. The background was determined by measuring
two metal film 1~k$\Omega$ resistors in bridge configuration.}
\label{Fig:Curves}
\end{figure}

Assuming that the $I_C R_N$-product is constant, we measure
fluctuations of the tunnel resistance $R_N$. This is in contrast to
most previous measurements, where the $I_C$ noise was measured
directly, but was shown to give the same relative noise
~\cite{Bib:Mueck} and also allows us to measure noise above $T_C$.
Our setup consists of two nominally identical samples in a bridge
configuration~\cite{Bib:Scofield} mounted in a $^3$He-cryostat with
a base temperature of about 250~mK (see Fig.~\ref{Fig:Setup} (a)).
The bias resistors have more than 1000 times the sample resistance
and are situated at room temperature. The samples are kept in the
normal state at all temperatures by applying a magnetic field well
beyond 100~mT and are biased with an ac current at 1 kHz, which is
the noise optimum of our amplifiers. We use a pair of home-made
battery-powered pre-amplifiers at room temperature with an input
voltage noise of less than $900\ \textrm{pV}/\sqrt{\textrm{Hz}}$,
followed by lock-in amplifiers. Their output signal is digitized and
the cross-correlation spectrum is computed to suppress amplifier and
cable noise. The noise floor of the system is determined by current
noise injected from the amplifier inputs to the sample (about
60~fA$/\sqrt{\textrm{Hz}}$), and residual microphonic pickup in the
voltage sensing cables. The long-term stability of the setup
permitted measurements down to 1~mHz at all temperatures (see
Fig.~\ref{Fig:Curves}). We determined the overall noise floor at a
typical source impedance by measuring two precision metal film
resistors of 1~k$\Omega$ at 300 mK.

The choice of bias currents was dictated by heating. For test
junctions in a geometry similar to flux qubits, the poor thermal
conductivity of the narrow leads limited us to 400~nA. Therefore, we
also fabricated samples with comparable junction areas, but rapidly
widening leads (see Fig.~\ref{Fig:Setup} (b)). This permitted bias
currents up to 2~$\mu$A. Since we suppressed superconductivity in
the Al lines, we could estimate the thermal conductivity seen from
the junction from the geometry and the measured sheet resistance
using the Wiedemann-Franz law~\cite{Bib:Pobell}. From this, we
calculated the self-heating to be below 100~mK at the highest bias
current and at the lowest temperature. We also repeated the
measurements at different excitation levels, to confirm that the
noise results were not affected by heating.

We prepared the samples using the double-angle shadow evaporation
method~\cite{Bib:Dolan}. The qubit test junctions were fabricated in
the same geometry and on the same chip as qubit devices with
junctions sizes between $140\times250\ \textrm{nm}^2$ and
$200\times250\ \textrm{nm}^2$. The samples with wide leads had areas
between $490\times190\ \textrm{nm}^2$ and $880\times190\
\textrm{nm}^2$ and were fabricated in a separate run using the same
evaporator.
\begin{figure}\includegraphics[width=8.5cm]{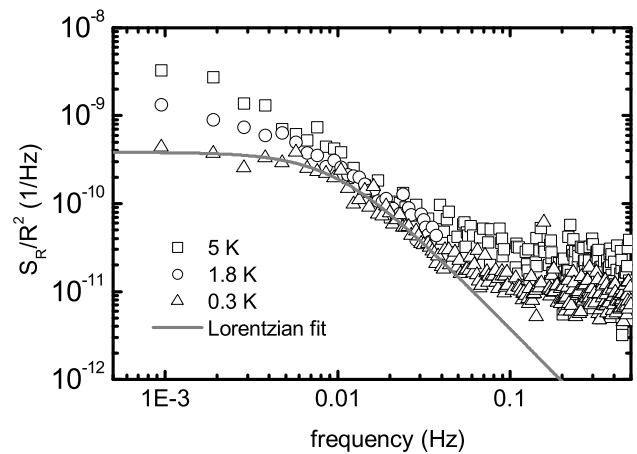}
\caption{Noise spectral density of the same sample as in Fig.~1, for
$T$ between 5~K and 0.3~K. At the lowest temperature, a Lorentzian
spectrum was fitted to the data.} \label{Fig:Lorentz}
\end{figure}

Figure~\ref{Fig:Curves} shows the resistance noise spectral density
of a pair of two tunnel junctions with $174\ \Omega$ and $185\
\Omega$ resistance and 0.17~$\mu $m$^2$ junction area each. The
results are normalized to the sample resistance, which facilitates
comparison between different samples. At room temperature, the
spectrum follows a $1/f$ law over the whole range from 1~mHz to
60~Hz. The spectral density decreases rapidly when cooling down.
Below 5~K the spectrum deviates from the $1/f$ behavior. This is
seen more clearly in Fig.~\ref{Fig:Lorentz}, where we plot the
normalized spectral density $S_R/R^2$ at low temperature. While the
$1/f$-like contribution keeps decreasing with $T$ at 1~mHz,
$S_R/R^2$ saturates around 10~mHz, where a Lorentzian shape
dominates the spectrum. Fitting Eq.~\ref{Eq:Lorentz} to the data in
Fig.~\ref{Fig:Lorentz} gave $\tau_{\mathrm{eff}} = 16$~s and $\delta
R = 0.85\ \textrm{m}\Omega$.
\begin{figure}\includegraphics[width=8.5cm]{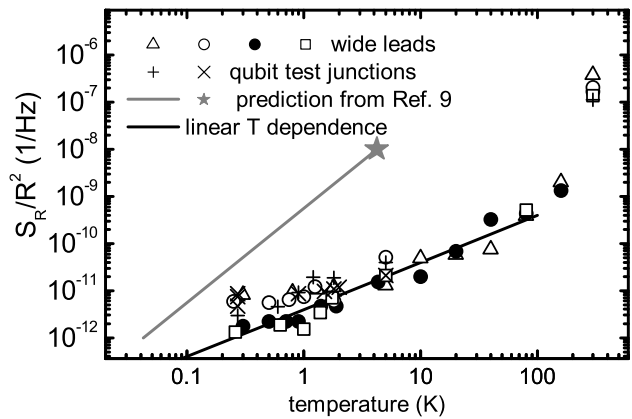}
\caption{Temperature dependence of the noise spectral density of
five different samples at $f=15$~mHz. The data was normalized to a
junction area of $1\ \mu \textrm{m}^2$. Open ($\circ$) and solid
($\bullet$) circles are data from the same sample taken in two
different cooldowns. The star ($\star$) is the average value from
Ref.~\onlinecite{Bib:vH} with the $T^2$-dependence (gray line)
assumed in that article. The solid black line is a fit to a linear
$T$-dependence.\label{Fig:TDep}}
\end{figure}

We observed a similar behavior in all the samples we measured. At
the lowest temperatures,  the spectrum consisted of a superposition
of a $1/f$-like background and one or two Lorentzians in the
observed frequency range. To compare different samples, we plot
$S_R/R^2$ at 15~mHz~\cite{Bib:Why15} of five junctions in
Fig.~\ref{Fig:TDep}, normalized to a junction area of $1\ \mu
\textrm{m}^2$ assuming a scaling of $S_R/R^2 \propto
1/\textrm{area}$~\cite{Bib:vH}. The data points for the different
samples are very close to each other for all temperatures. In fact,
the variation between different samples is about as large as between
different cooldowns (see below). We also plotted the noise
prediction from Ref.~\onlinecite{Bib:vH}. It exceeds our data by
more than two orders of magnitude at 4.2~K~\cite{Bib:IvsRNoise}.
Also, the observed temperature dependence deviates considerably from
the proposed $T^2$-behavior. From room temperature to 150~K,
$S_R/R^2$ decreases by about a factor of 100, and then follows a
linear $T$-dependence down to about 1~K. From the data points for
individual samples at lower temperatures we find that the noise
spectral density always saturates below a temperature around 0.8~K.
The saturation  was not due to a lack of sensitivity at lower $T$
(cf. Figs.~\ref{Fig:Curves}, \ref{Fig:Lorentz}) nor to heating, but
rather due to the saturation of individual defects dominating the
spectrum, as exemplified in Fig.~\ref{Fig:Lorentz}. If many of those
defects are present, their spectra are superimposed and lead to
$1/f$-noise with a linear~\cite{Bib:DH} or
quadratic~\cite{Bib:Wellstood,Bib:Shnirman} temperature scaling,
depending on the details of the defect dynamics. This picture breaks
down if a small number of fluctuators dominate the spectrum. At low
temperatures, we assume that their dynamics changes from the thermal
to the quantum regime and the temperature dependence saturates. In
this picture, we can assume that the noise level will not decrease
any further when going to below 300~mK. At 100~mK, where qubit
dephasing times were calculated in Ref.~\onlinecite{Bib:vH}, the
observed noise spectral density is again comparable to their
estimate, albeit due to a completely different temperature
dependence.

We also evaluated the time constants and resistance change caused by
the two-level fluctuators. The time constants, between 1\ s and 60\
s, were determined by the frequency window of our measurements.
Assuming that the current distribution in the junctions is
homogeneous and that a defect can block current flow completely in a
small part of the junction, we obtain effective defect areas of
typically 1\ nm$^2$ to 2\ nm$^2$.

The two-level fluctuators were changing with thermal cycling. For
example, the open and solid circles in Fig.~\ref{Fig:TDep} are
measurements of the same sample in two consecutive cooldowns. We
observed a Lorentzian superimposed on a $1/f$ background in the
first cooldown ($\circ$), and only the $1/f$ spectrum in the second
cooldown ($\bullet$). This leads us to assume that the lowest data
points in our graph are actually the closest estimate for $1/f$
noise, yielding $1.3\times 10^{-12}/\textrm{Hz}$ at 15~mHz. This
works out to $2.0\times 10^{-14}/\textrm{Hz}$ at 1~Hz, if we assume
a true $1/f$ dependence and normalize to 1~$\mu\textrm{m}^2$. Note
that at 1~Hz and at the lowest temperatures the $1/f$-noise
approaches the sensitivity of our setup.

For qubit decoherence, we have to distinguish between $1/f$-noise
and individual bistable fluctuators. In the former case, due to the
divergence as $f\rightarrow 0$, the total time for an experiment as
well as the details of signal averaging are important, as pointed
out in Ref.~\onlinecite{Bib:vH}. Using spin-echo, the coherence time
can be improved only by a logarithmic factor~\cite{Bib:Ithier} of
the order five. Random telegraph noise in the barrier resistance, on
the other hand, leads to a splitting of the qubit spectroscopy lines
and beats in the coherent oscillations, if the qubit energy
landscape switches frequently during an experimental run. Unlike
$1/f$-noise, it falls off quickly at high frequencies and can
therefore be efficiently suppressed with spin-echo pulses or more
advanced pulse sequences, such as series of short
$\pi$-pulses~\cite{Bib:BB}. Dephasing times due to pure $1/f$-noise
were estimated in Ref.~\onlinecite{Bib:vH} at $T=100$~mK, which gave
a normalized spectral density of a 1~$\mu \textrm{m}^2$ junction of
$8.2\times 10^{-14}/\textrm{Hz}$ at 1 Hz. In our junctions, we
observed a $1/f$-noise of about a quarter of that value at 300~mK,
and we assume that the noise has already saturated at that
temperature. If we suppose that the $I_C$-noise in qubit junctions
is only due to $R$-noise we get twice the dephasing times estimated
in Ref.~\onlinecite{Bib:vH}. For the three-junction flux qubit,
charge-phase and phase qubits, this works out to 1.6~$\mu$s,
3.6~$\mu$s, and 28~$\mu$s, respectively. As an example, the
spin-echo time of 4~$\mu$s in a recent flux qubit
experiment~\cite{Bib:Patrice4us} was still limited by the
measurement circuitry, but already approached the expected spin-echo
time due to $1/f$-noise ($\approx 5\times 1.6\ \mu\textrm{s}$).

To summarize, we have measured low-frequency resistance fluctuations
in aluminum based Josephson junctions, as used for superconducting
qubits. The noise spectral density at 4.2~K is two orders of
magnitude lower than expected from the literature survey in
Ref.~\onlinecite{Bib:vH}, and we find a linear $T$-dependence
between 150~K and 1~K instead of the proposed $T^2$-law. The
spectral density saturates below 0.8~K, which is due to individual
strong two-level fluctuators. The dephasing times due to pure
$1/f$-noise are estimated to be about twice as long as in
Ref.~\onlinecite{Bib:vH}.

We would like to thank P. Bertet, A. Lupa\c{s}cu and R. Simmonds for
discussions. Financial support by FOM, NanoNed and the EU through
the projects SQUBIT-2 and EuroSQIP is gratefully acknowledged.


\begin{thebibliography}
\bibitem{Bib:IvsRNoise}
This comparison assumes that $I_C$-noise is due to resistance
fluctuations only. Although this is generally believed to be correct
(see also Ref.~\onlinecite{Bib:Mueck}), we cannot exclude additional
noise sources contributing only to $I_C$-noise, but not to
$R$-noise.
\end{thebibliography}
\end{document}